\documentclass[aps,twocolumn]{revtex4}
\newcommand*{\be}{\begin{equation}} 
\newcommand*{\ee}{\end{equation}}
\newcommand*{\bqn}{\begin{eqnarray}}
\newcommand*{\eqn}{\end{eqnarray}}

\usepackage{graphicx}
\usepackage{color}

\begin{document}

\title{Uncertainty principle in a cavity at finite temperature }
\author{A. P. C. Malbouisson$^{a}$ }
\address{(a) {\it Centro
Brasileiro de Pesquisas F\'{\i}sicas/MCTI, 22290-180, Rio de
Janeiro, RJ, Brazil}}
\email[E-mail address: ]{adolfo@cbpf.br (A.P.C. Malbouisson)}

\begin{abstract}

We employ a dressed state approach to perform a study on the behavior of the uncertainty principle for a system in a heated cavity. We find, in a small cavity for a given temperature, an oscillatory behavior of the momentum--coordinate product, $(\Delta\,p)\,(\Delta\,q)$, which attains periodically finite absolute minimum (maximum) values, no matter 
large is the elapsed time.  This behavior is in a sharp contrast with 
what happens in free space, in which case, the product $(\Delta\,p)\,(\Delta\,q)$ tends asymptotically,  for each temperature,  to a constant  value, independent of time.

\vspace{0.34cm}
\noindent

PACS number(s): 03.65.Ta 

\end{abstract}
\maketitle

{\it Introduction}

An account on the subject  of an interacting particle-environment system, 
 can be found in Refs.~\cite{zurek,paz,haake,caldeira,livro}, in which the environment 
 is considered as an infinite set of noninteractng oscilllators. 
 Here we consider a similar model, treated with a different approach. 
  Let us therefore begin with some words about the method we employ: From a general point of view, 
 apart from computer calculations in lattice field theory, the must currently used  
 method to treat the physics of interacting particles is perturbation theory, in which 
 the starting point are bare fields (particles) interacting by means of gauge fields.  
 Actually, as a matter of principle,  the idea of 
 bare particles associated to  bare matter fields and of a  gauge particle mediating the 
interaction among them, is in fact an artifact of perturbation
theory and, strictly speaking, is physically meaningless. A charged physical particle is always coupled to 
the gauge field,  it is always "dressed" by a cloud of quanta of the gauge
field (photons, in the case of Electrodynamics). Exactly the same type of argument 
 applies {\it mutatis} {\it mutandis} to a  particle-environment system, 
 in which case we may speak of a "dressing"
of the  particle by the thermal bath, the 
 particle being "dressed" by a cloud of quanta of the environment. 
This should be true in general for any system in which a material particle is coupled to a  
field, no matter the specific nature of the field (environment) nor of the interaction involved.
 We give a treatment to this kind of system using some {\it dressed} (or renormalized) 
coordinates.  In terms of these new coordinates {\it dressed} states are defined, which 
allow to divide the coupled system into two parts, the {\it dressed}
particle and the {\it dressed} environment, which makes unnecessary to work 
directly with the concepts of bare particle, bare environment and interaction 
between them.  A detailed exposition of our formalism and of its meaning, for both zero- and 
finite temperature can be found in Refs.~\cite{linhares,termalizacao,termica,emaranho,erico}. 

About the physical situation we deal with, on  general grounds, very precise investigations have been done 
on the fundamentals of quantum physics, in particular on the validity of the Heisenberg uncertainty relation. In~\cite{clerk}, 
it is reported  that a great deal of 
effort is being made  to minimize  external noising factors, 
 such as thermal fluctuations and electricity oscillations in experiments,  
in order to verify the  relation, in the spirit of zero-temperature quantum physics. 
However, changes in the uncertainty principle induced by temperature is an idea already explored  in the literature, in particular for open systems. In~\cite{Abe}, the authors study with a thermo-field-dynamics formalism, the relation between the sum of information-theoretic entropies in quantum mechanics with measurements of the position and momentum of a particle surrounded by a thermal environment. It is found that this quantity cannot be made arbitrarily small but has a universal lower bound dependent on the temperature.  They also show that the Heisenberg uncertainty relation at finite temperature can be derived in this context. In~\cite{Hu,Hu1} it is obtained the uncertainty relation for a quantum open system consisting of a 
Brownian particle interacting with an 
ohmic bath of quantum oscillators at finite temperature. 
These authors claim that this allows to get  some insight into the physical mechanisms involved in the environment-induced decoherence process. 
Also, modifications of the uncertainty principle have been proposed in, for instance, 
a cosmological context. As remarked in~\cite{amelino}, in quantum gravity it seems to be 
needed a  generalized position--momentum uncertainty principle. The authors of 
 ref.~\cite{amelino} investigate a possible connection between the generalized  uncertainty principle and  
 changes in the area--entropy black hole formula and the black hole evaporation process. 

In this report, we study the behavior of the particle-environment system contained 
in a cavity of arbitrary size, under the influence of a heated environment. The environment is 
composed of an infinity of oscillators, and assumed to be at a given temperature, realized by taking  
an appropriate thermal distribution for its modes. 
This generalizes previous works for zero temperature, as for instance in Refs.~\cite{adolfo3,adolfo4,adolfo5},  
for both inhibition of spontaneous decay in cavities and the Brownian motion. 
We study the time dependent mean value for the dressed 
oscillator position operator  taken in a {\it dressed} coherent state. We find that in both cases,
of the environment at zero temperature or of the heated environment, 
these mean values are independent of the temperature and are given by the same expression.  
On the other side,  the mean squared error for both the particle position and momentum 
 are dependent on the temperature. From them we get the time and 
 temperature-dependent uncertainty Heisenberg 
 relation.  We then investigate how heating 
affects the uncertainty  principle in a cavity of arbitrary size. This is particularly 
interesting in a small cavity, where the result is not a trivially expected one.

{\it The model}

Our approach to the problem
makes use of the notion of \textit{dressed thermal states}~\cite{termica},
in the context of a model already employed in the literature, of atoms, or
more generally material particles, in the harmonic approximation, coupled to
an environment modeled by an infinite set of point-like harmonic oscillators
(the field modes). The dressed thermal state approach is an extension of the
dressed (zero-temperature) formalism already used earlier~\cite{adolfo1,adolfo3,adolfo4,adolfo5}. 

We  consider a bare particle (atom, molecule,...) approximated by a harmonic
oscillator described by the bare coordinate and momentum $q_0, p_0$
respectively, having \emph{bare} frequency $\omega _0$, linearly
coupled to a set of $N$ other harmonic oscillators (the environment)
described by bare coordinate and momenta $q_k, p_k$ respectively,
with frequencies $\omega _k$, $k=1,2,\ldots ,N$. The limit
$N\rightarrow \infty$ will be  undesrstood.  The whole system is
supposed to reside inside a perfectly reflecting spherical cavity of radius $R$ in
thermal equilibrium with the environment, at a temperature
$T=\beta^{-1}$. The system is described by the Hamiltonian 
\begin{equation}
H=\frac 12\left[ p_0^2+\omega _0^2q_0^2+\sum_{k=1}^N\left(
p_k^2+\omega _k^2q_k^2\right) \right] - q_0\sum_{k=1}^Nc_kq_k,
\label{Hamiltoniana}
\end{equation}
The Hamiltonian (\ref{Hamiltoniana}) is transformed to principal
axis by means of a point transformation, 
$q_\mu  =\sum_{r=0}^{N}t_\mu ^rQ_r\,,\,\,\,p_\mu =\sum_{r=0}^{N}t_\mu ^rP_r;\;
\mu  =(0,\{k\}),\; k=1,2,...,N;\;\;r=0,...N$,  
performed by an orthonormal matrix $T=(t_\mu ^r)$. The subscript
 $r$ refers to the normal
modes. In terms of normal momenta and coordinates, the transformed
Hamiltonian  reads
$H=\frac 12\sum_{r=0}^N(P_r^2+\Omega _r^2Q_r^2)$,  
where the $\Omega _r$'s are the normal frequencies corresponding
to the collective \textit{stable} oscillation modes of the coupled
system. Using the coordinate transformation in the
equations of motion  and explicitly making use of the
normalization of the matrix $(t_{\mu}^{r})$, $\sum_{\mu =0}^N(t_\mu ^r)^2=1$,    
 we get the matrix elements $(t_{\mu}^{r})$~\cite{adolfo1}.  

We take $c_k=\eta (\omega _k)^u$, where $\eta $ is a constant
independent of $k$. In this case the environment is classified
according to $u>1$, $u=1$, or $u<1$, respectively as
\textit{supraohmic}, \textit{ohmic} or \textit{subohmic}~\cite{paz,haake}; 
we take, as in~\cite{adolfo1}, $\eta=2\sqrt{g\Delta \omega /\pi }$, 
where $\Delta \omega$ is the interval between two 
neighboring field frequencies and $g$ a fixed constant 
characterizing the strenght  of the coupling particle-environment. 
 Restricting ourselves to an ohmic  environment, we get an equation for the 
 $N+1$ eigenfrequencies $\Omega _r$, 
corresponding to the
$N+1$ normal collective modes~\cite{adolfo1}. In this 
  case the eigenfrequencies 
  equation contains a divergence for $N\rightarrow
\infty $ and a 
renormalization procedure is needed. This leads to 
 the \textit{renormalized} frequency~\cite{adolfo1} (this renormalization procedure was pioneered in~\cite{Thirring}), 
\begin{equation}
\bar{\omega}^2=\omega _0^2-\delta \omega ^2=\lim _{N \rightarrow
\infty }(\omega_{0}^2 - N\eta^2). \label{omegabarra}
\end{equation}
where we have defined the counterterm
$\delta \omega ^2=N\eta ^2$.  


We introduce {\it dressed} or \textit{renormalized}  coordinates $q_0^{\prime }$ and
$\{q_k^{\prime }\}$ for, respectively, the \textit{dressed } atom
and the \textit{dressed} field, defined by,
\begin{equation}
\sqrt{\bar{\omega}_\mu }q_\mu ^{\prime }=\sum_rt_\mu
^r\sqrt{\Omega _r}Q_r, \label{qvestidas1}
\end{equation}
 where $\bar{\omega}_\mu
=\{\bar{\omega},\;\omega _k\}$. In terms of these, we
define thermal  \textit {dressed} states, precisely described in~\cite{termalizacao,termica}. 

It is worthwhile to note that our renormalized coordinates  are
objects different from both the bare coordinates, $q$, and the
normal coordinates $Q$. Also, our dressed  states, although being
collective objects, should not be confused with the eigenstates
 of the system~\cite{footnote}. 
In terms of our renormalized coordinates and dressed states, we can find  a natural
division of the system into the dressed (physically observed) particle and
 the  dressed environment. The dressed particle  will contain automatically
all the effects of the environment on it.

{\it A cavity of arbitrary size at finite temperature}

To study the behavior of the system in a cavity of arbitrary size,  we  write the initial
physical state in terms of dressed coordinates. We assume that initially the system is described by the density operator,
$\hat{\rho}(0)=\hat{\rho}_0´\otimes \hat{\rho}_\beta'\;,$
where $\hat\rho_0´$ is the density operator associated with the oscilator $q_0'$, that
can be in general in a pure or mixed state.  Also, 
$\hat{\rho}_\beta'$ is the dressed density operator associated with the dressed field
modes. We assume thermal equilibrium for these dressed modes, thus
\begin{equation}
 \hat{\rho}_\beta'=\frac{\bigotimes_k e^{-\beta \hat{H}_k'}}{{\rm Tr}\bigotimes_k e^{-\beta \hat{H}_k'}}\,\;{\rm{with}}\;\hat{H}_k'=\frac{1}{2}\hat{p}_k'^2+\frac{1}{2}\omega_k^2\hat{q}_k'^2\;.
\label{eqd2}
\end{equation}
The time evolution of the density operator is given by the Liouville-von Newman equation, whose solution,
in the case of an entropic evolution, is given by
$\hat{\rho}(t)=e^{-\frac{i}{\hbar}\hat{H}t}\hat{\rho}(0)e^{\frac{i}{\hbar}\hat{H}t}\;.$
Then, the time evolution of the average thermal expectation value of an operator is given by, 
\begin{equation}
\langle \hat{A}\rangle(t)={\rm Tr}\left(\hat{A}e^{-\frac{i}{\hbar}\hat{H}t}\hat{\rho}(0)
e^{\frac{i}{\hbar}\hat{H}t}\right)\,={\rm Tr}\left(\hat{A}(t)\hat{\rho}(0)\right)\,
\end{equation} 
where the ciclic property of the trace has been used; above, 
$\hat{A}(t)=e^{\frac{i}{\hbar}\hat{H}t}\hat{A}e^{-\frac{i}{\hbar}\hat{H}t}$ is the time-dependent operator 
$\hat{A}$ in the 
Heisenberg representation. 


{\it Dressed coherent states in a heated environment}

Let us consider a  Brownian 
particle embedded in a heated environment as described above.  In our language, 
we speak of a dressed Brownian 
particle and we  use the {\it dressed} state formalism.  
We assume, as usual, that initially the  particle and the environment are
decoupled and that the coupling is turned on suddenly at some given time, that we choose at
$t=0$. 
In our formalism, we define  
$|\lambda\left.\right>$ as a {\it dressed} coherent state given  by, 
\be
|\left. \lambda, n_1',n_2',..;t=0\right>={\rm e}^{-|\lambda|^2/2}
\sum_{n_0'=0}^{\infty}\frac{\lambda^{n_0'}}
{\sqrt{n_0'!}}|\left.n_0'n_1',..\right>\;, 
\label{eq14}
\ee
where $n_0^{\prime}$ stands for the occupation number of the dressed particle 
and where $n_1',n_2',..$ are the occupation numbers of the field modes. For 
zero temperature we have $n_1'=n_2',..=0$. 
For finite temperature,  
we will do computations  taking  $\hat{\rho}_0=|\lambda\rangle\langle\lambda|$. This means,
that at the initial time, the dressed particle oscillator is in a pure coherent state.
Keeping this in mind, we consider the quantity, 
 $\langle \hat{q}_0'\rangle(t)$, that we denote by $q_0'(t)$, 
$q_0'(t)={\rm Tr}\left(\hat{q}'_0(t)\hat\rho(0)\right)$.
In order to evaluate the  above expression we first compute $\hat{q}'_0(t)$. Using  the relation
between the dressed coordinates and the normal coordinates, Eq.(\ref{qvestidas1}), 
 the  expression for $\hat{H}$ in terms of the normal coordinates, 
and the Baker-Campbell-Hausdorff formulas, we
get, after some steps of calculation, 
\begin{equation}
 q_0'(t)=\sqrt{\frac{\hbar}{2\bar{w}_0}}\left(  \lambda f_{00}(t)+\lambda^\ast f_{00}^\ast(t) \right), 
\label{eqd9}
\end{equation} 
where $f_{00}(t)$ is one of the quantities $f_{\mu \nu }(t) =\sum_s t_\mu ^st_\nu ^se^{-i\Omega _st}$~\cite{adolfo3}.
 
Note that the above expression is independent of the temperature and coincides with the one
obtained previously for the the zero temperature case~\cite{adolfo3}. This is because 
$\hat{\rho}_\beta$ has even parity  in the dressed momentum and position  operators. 
Due to the same reason, we find an entirely similar 
formula for $ p_0'(t)$.
The situation is different for the quantity $q_0'^2(t)=\langle  \hat{q}_0'^2\rangle(t)$. 
After performing similar computations as above,
we get,
\begin{eqnarray}
q_0'^2(t,\beta)=\frac{\hbar}{2\bar{\omega}}\left[  \left(\lambda f_{00}(t)+\lambda^\ast f_{00}^\ast(t)\right)^2 \right. \nonumber \\
\left. +2\sum_k | f_{0k}(t)|^2n_k'(\beta)+1\right]\;,
\label{eqd10}
\end{eqnarray}
Then from Eqs. (\ref{eqd9}) and (\ref{eqd10}) we obtain for the mean square error, 
\begin{eqnarray}
(\Delta q'_0)^2(t,\beta)&=&\langle \hat{q}'^2_0\rangle(t,\beta) -(\langle\hat{q}_0'\rangle(t))^2
\nonumber\\
&=&
\frac{\hbar}{2\bar{\omega}}+ \frac{\hbar}{\bar{\omega}}\sum_k| f_{0k}(t)|^2n_k'(\beta)\;,
\label{eqd11}
\end{eqnarray}
where  $n_k'(\beta)$ is given by the Bose-Einstein distribution,
$n_k'(\beta)=1/(e^{\hbar\beta\omega_k}-1)$~\cite{termica}.

Analogously we obtain  
 the momentum mean squared error, 
\begin{eqnarray}
(\Delta p_0)^2(t,\beta)&=& p_0'^2(t,\beta)-(p_0'(t))^2\nonumber\\
&=&\frac{\hbar\bar{\omega}}{2}+\hbar \bar{\omega}\sum_k|f_{0k}(t)|^2n_k'(\beta).   
\label{eqd15}
\end{eqnarray}
From Eqs. (\ref{eqd11}) and (\ref{eqd15}) we obtain the time and temperature-dependent Heisenberg relation,
\begin{equation}
\Delta q_0'(t,\beta)\Delta p_0'(t,\beta)=\frac{\hbar}{2}+\hbar\sum_k |f_{0k}(t)|^2n_k'(\beta)\;.
\label{eqd16}
\end{equation}

{\it Time behavior in a cavity with a heated environment}


Let us consider the time evolution of the uncertainty relation $\Delta q_0'(t,\beta)\Delta p_0'(t,\beta)$ 
given in Eq.~(\ref{eqd16}), in a finite (small) cavity, characterized by  the dimensionless 
parameter $\delta =gR/\pi c$ and  take a 
coupling regime defined by a relation between $g$ and the 
emission frequency $\bar{\omega}$, $g=\alpha \,\bar{\omega}$ . For instance, if we consider $\delta=0.1$,  
$\alpha =0.2$ and  $\bar{\omega} \approx 10^{14}/{\rm{s}}$ (in the visible red), 
this corresponds to a cavity radius $R\sim 10^{-6}{\rm{m}}$. We 
measure the uncertainty  relation in units of $\hbar$ and call it for simplicity 
$\Delta q_0'(t,\beta)\Delta p_0'(t,\beta)=\Delta(t,\beta)$. 
Then calculations can be performed in a similar 
way as in Ref.~\cite{termica} and we obtain (there is no confusion between the variable $t$ describing time 
and the matrix elements $t_{\mu}^{r}$),  
\begin{eqnarray}
\Delta(t,\beta) =\frac{1}{2}+
\sum_{k=1}^{\infty}\frac{1}{e^{(\hbar\beta g/\delta)k}-1}\left[(t_0^0)^2(t_k^0)^2 \right.\nonumber \\
\left.+2\sum_{l=1}^{\infty}t_0^0t_0^lt_k^0t_k^l\cos(\Omega_0-\Omega_l)t \right.\nonumber \\
\left.+ \sum_{l,n=1}^{\infty} t_0^lt_0^nt_k^lt_k^n
\cos(\Omega_l-\Omega_n)t \right]\,.\nonumber \\
\label{n0tbeta}
\end{eqnarray}
 The matrix
elements $t_{\mu}^{r}$  in the above formulas are
evaluated in~\cite{termica}, 
\begin{eqnarray}
t_{k}^{0} \approx \frac{k\,g^2\sqrt{2\delta}}{k^2g^2-\Omega_0^2\delta^2}\,;\;\;
t_{k}^{l} \approx \frac{2k\delta}{k^2-(l+\epsilon_{l})^2}\frac{1}{l} \nonumber \\
(t_0^k)^2 \approx \frac{2gR}{\pi c k^2} = \frac{2 \delta}{k^2} \,,\;\;
(t_0^0)^2 \approx 1-\frac{\pi g R}{3c} = 1 - \frac{\pi^2 \delta}{3},
\nonumber \\
\label{ts}
\end{eqnarray}
where $\epsilon_{l}$ is a small quantity such that $0<\epsilon_{l}<1$. Actually, for a 
small cavity ($\delta \ll 1$) $\epsilon_{l}\approx \delta/k$.

\begin{figure}[t]
\includegraphics[{height=6.0cm,width=8.5cm,angle=360}]{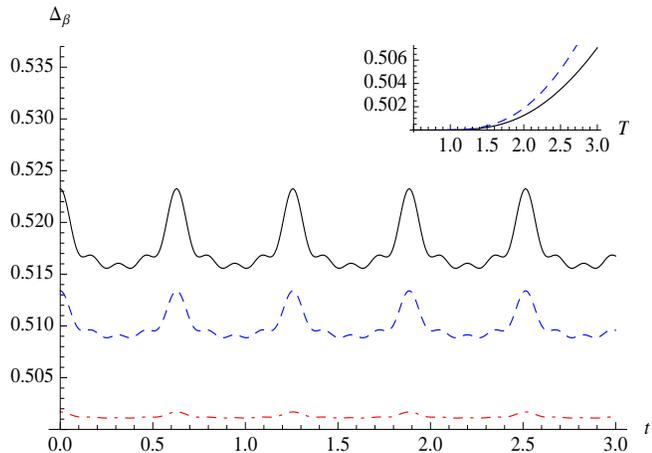}
\caption{Time evolution of the thermal dependent uncertainty relation 
$\Delta_{\beta}(t)$ for three different values of the temperature $T=3.85\,(\beta=0.26)$ (full line),  $T=3.57\,(\beta=0.28)$ (dashed line) and $T=1.96\,(\beta=0.51)$ (dotdashed line), from the upper to the lower curves, respectively .  We take $g=1.0$, $\bar{\omega}=5.0$ and $\delta =0.1$; the scale for the temperature, the vertical axis and for time is 
 in units such that $k_B=\hbar =c=1$;  in the upper right small figure it is shown the temperature dependence 
 of two neighboring minimum and maximum values of  $\Delta_{\beta}(t)$, respectively full and dashed lines,
 for two times, $t\approx 2.3$ and $t\approx 2.5$, where they occur} \label{FigDelta1}
\end{figure}
{\it Comments}

 Eq.~(\ref{n0tbeta}) 
describes  the time evolution of  the uncertainty relation 
$\Delta q_0'(t,\beta)\Delta p_0'(t,\beta)\equiv \Delta(t,\beta)$ in a small cavity. 
A plot of this time evolution is given in Fig.~\ref{FigDelta1} 
for some representative values of the temperature. 
The thermal uncertainty function, $\Delta(t,\beta)$, is an oscillating
function which attains periodically an absolute minimum (maximum) value,
${\rm{Min}}[\Delta(t,\beta)]$ (${\rm{Max}}[\Delta(t,\beta)]$).  Since 
the periodical character of $ \Delta(t,\beta)$ does not involve $\beta$, the 
location  of these extrema on the time axis 
do not depend on the temperature. Indeed, we can see from Fig.\ref{FigDelta1}, that  
the values of these minima and maxima  depend on the temperature,  
$\beta^{-1}$, but appears to be, for each temperature, the same for all values of time where they occur,
in other words the values of the absolute extrema appears to be independent of time. 
 In the detail of Fig.\ref{FigDelta1}, are plotted two neighboring absolute minimum and 
maximum (corresponding to $t\approx 2.3$ and $t\approx 2.5$), as  functions of temperature.  
We find from these figures that raising the temperature increases the
amplitude of oscillation and the mean value of the uncertainty relation and that its
 lower and upper bounds, also grows with temperature. 
 
 We infer from Fig.\ref{FigDelta1} that for a small cavity, in all cases an oscillatory behavior is 
 present for $\Delta(t,\beta)$, with the amplitude of the oscillation depending on the temperature, $T$.
  For larger values of $T$ the 
 amplitude of the oscillation, and both, its absolute minimum and maximum values, 
 are larger than for lower temperatures. 
 This behavior of the uncertainty principle  is to be contrasted with 
the case of an arbitrarily large cavity (free space). In this last case,  the product  $(\Delta\,p)\,(\Delta\,q)$ 
goes, asymptotically, as $t\rightarrow \infty$ for each temperature,  to a constant  value $\Delta(\beta)$. 
This asymptotic value depends on the temperature and grows with it, but is independent of time.
Distinctly, for a small cavity, even for $t\rightarrow \infty$, the product $(\Delta\,p)\,(\Delta\,q)$ presents 
oscillations, which have  larger and larger amplitudes for higher temperatures.  

The result above falls in a general context of the different behaviors of quantum systems confined in 
cavities, as compared to free space, in both, zero or finite temperature, the system 
being investigated using the formalism presented in this report.  
In~\cite{adolfo3} some of 
us got the expected result 
that the probability, $P(t)$,  that a simple cold atom in free space, excited at $t=0$ remain excited 
after an elapsed time $t$, decays monotonically, going to zero as $t\rightarrow \infty$, while  
in a small cavity, 
$P(t)$ has an oscillatory behavior, never reaching zero. For $\bar{\omega}\sim 10^{14}$ (in the visible red), 
$R\approx 10^{-6}{\rm{m}}$, in a weak (of the order of electromagnetic) coupling regime, ${\rm{Min}}\,P(t)\approx 0.98$, 
which is in agreement with experimental observations~\cite{Haroche3}. At finite temperature it is obtained in~\cite{termica}, 
for a small cavity,  that the occupation number of a simple atom in a heated environment has an oscillatory 
behavior with time and that its mean value increases with increasing temperature. In~\cite{emaranho,erico} the 
behavior of an entangled bipartite system at zero and finite temperature is investigated; taking two measures of 
entanglement, it results an oscillatory behavior for a small cavity (entanglement is preserved at all times), 
while it disappears as $t\rightarrow \infty$ for free space. 

 We hope that the result presented in this report could have some usefulness in nanophysics or in quantum 
 information theory. At this moment we are not able to comment about these aspects; they will be subject of future studies.

{\bf Acknowledgements}: The author thanks FAPERJ and CNPq (brazilian agencies) for partial financial support.



\begin{thebibliography}{99}
\bibitem{footnote} Notice that  our dressed states  are not the same as those employed 
in studies involving the interaction of atoms and
electromagnetic fields~\cite{bouquinCohen} and in the study of the radiation damping of
classical systems~\cite{petrosky}.
\end{thebibliography}
\end{document}